\documentclass[journal]{IEEEtran}
\usepackage{amsmath,amsfonts}
\usepackage{algorithmic}
\usepackage{algorithm}
\usepackage{array}
\usepackage[caption=false,font=normalsize,labelfont=sf,textfont=sf]{subfig}
\usepackage{textcomp}
\usepackage{stfloats}
\usepackage{url}
\usepackage{verbatim}
\usepackage{graphicx}
\usepackage{cite}
\begin{document}

\title{Reliability-Constrained Hybrid Beamforming for Multistatic ISAC in Vehicular Networks}
\author
{
{Congcong Liu, Junhui Zhao,~\IEEEmembership{Senior Member,~IEEE}, Xiaoming Wang, Dongming Wang}
\thanks
{
This work was supported by the Fundamental Research Funds for the central Universities (2025JBZX060).

Congcong Liu, and Junhui Zhao are with the School of Electronic and Information Engineering, Beijing jiaotong University, Beijing 100044, China (e-mail: 21111041@bjtu.edu.cn; junhuizhao@hotmail.com).

Xiaoming Wang is with the College of Telecommunications and Information Engineering, Nanjing University of Posts and Telecommunications, Nanjing 210003, China (e-mail: xmwang@njupt.edu.cn). 

Dongming Wang is with the National Mobile Communications Research Laboratory, Southeast University, Nanjing 210096, China (e-mail: wangdm@seu.edu.cn).
}
}
\maketitle

\begin{abstract}
This letter investigates reliability-constrained hybrid beamforming for transceiver-separated multistatic integrated sensing and communication in vehicular networks. A target-position Cramér--Rao bound  minimization problem is formulated under outage-probability, transmit-power, and analog constant-modulus constraints. To handle the constrained non-convex problem, we develop a proportional-integral Lagrangian proximal policy optimization algorithm. 
Simulation results show that the proposed algorithm keeps the average outage probability at or below the reliability threshold, around 8\%--10\%, improves constraint satisfaction, and achieves stable sensing performance.

\end{abstract}


\section{Introduction}

\IEEEPARstart{I}{ntegrated} sensing and communication (ISAC) is regarded as a promising candidate technology for sixth-generation mobile communication systems \cite{RenIoV,9585321}. 
By sharing spectrum, hardware, and signal waveforms, ISAC enables wireless communication and environmental sensing over a unified platform \cite{RenITS}. In vehicular networks, applications such as autonomous driving, cooperative sensing, blind-spot warning, and road environment monitoring require both reliable data transmission and real-time sensing capability \cite{9830717,10543024}. However, high vehicle mobility, frequent road obstructions, and dynamic service demands make ISAC resource allocation and beamforming design challenging \cite{10063187}.

Communication and sensing tasks in vehicular ISAC systems share limited spectrum, power, and spatial beam resources, but their performance objectives are different \cite{9945983}. Communication focuses on link reliability and transmission rate, whereas sensing requires sufficient target-direction beam gain and estimation accuracy. Steering beams toward vehicular users can improve communication reliability but may weaken target illumination, while sensing-oriented beams may degrade vehicular links. Therefore, reliability-constrained beamforming is essential for balancing communication and sensing performance in vehicular ISAC.

Existing studies on ISAC beamforming mainly focus on monostatic architectures, where the same node performs signal transmission and target echo reception \cite{11370420,RenRIS}. Although simple to deploy, such architectures may suffer from transmit-receive coupling interference, restricted observation angles, and unfavorable reflection links in complex road environments. In contrast, transceiver-separated multistatic ISAC provides a feasible alternative, where the active access point transmits downlink signals and illuminates the target, while multiple passive access points receive target echoes from different perspectives \cite{11153961,11026097}.

Motivated by the above, this letter investigates a reliability-constrained hybrid beamforming problem for transceiver-separated multistatic ISAC in vehicular networks. Specifically, we consider a system consisting of one active access point, multiple passive access points, multiple vehicular users, and one point target. The active access point transmits downlink ISAC signals to serve vehicular users and illuminate the target, while the passive access points cooperatively receive target echoes for multistatic sensing. The objective is to improve target localization performance under outage-probability, transmit-power, and analog constant-modulus constraints. To handle the coupled communication reliability and sensing performance, as well as the continuous non-convex beamforming variables, we develop a reliability-aware beamforming optimization method.

The  contributions of this letter are summarized as follows:

\begin{enumerate}
\item We construct a transceiver-separated multistatic ISAC model, where  active access point serves vehicular users and multiple passive access points receive target echoes.

\item We formulate a reliability-constrained hybrid beamforming problem to minimize the point-target CRB under outage-probability, transmit-power, and analog constant-modulus constraints.

\item  We develop a proportional-integral Lagrangian proximal policy optimization based method for feasible online beamforming. Simulations show that the proposed method achieves stable convergence, and higher constraint satisfaction than baseline methods.

\end{enumerate}

\section{System Model}

We consider a transceiver-separated multistatic ISAC system for vehicular networks, consisting of one active AP, multiple passive APs, $K$ downlink vehicular users, $U$ uplink users, and one point target. The active AP transmits ISAC signals to serve downlink users and illuminate the target, while the passive APs cooperatively receive target echoes from different spatial perspectives. The uplink users are regarded as external interference sources, and their transmit powers are treated as given parameters rather than optimization variables.

The active AP is equipped with $M_t$ transmit antennas and $M_{\mathrm{RF}}$ RF chains, and adopts a hybrid beamforming architecture. Each passive AP has $M_r$ receive antennas, and all vehicular users are equipped with a single antenna. 
Let $\mathcal{K}=\{1,2,\ldots,K\}$, $\mathcal{U}=\{1,2,\ldots,U\}$, and $\mathcal{Z}_p=\{2,3,\ldots,Z\}$ denote the sets of downlink users, interfering uplink users, and passive APs, respectively.

\subsection{Signal Model}

At time slot $t$, the active AP transmits the communication symbol vector 
$\mathbf{x}[t]\in\mathbb{C}^{K\times 1}$ to $K$ downlink vehicular users. Following the ISAC signal reuse principle, the communication symbols are also reused as sensing probing waveforms without introducing dedicated sensing signals.

Let $\mathbf{W}_{\mathrm{RF}}[t]\in\mathbb{C}^{M_t\times M_{\mathrm{RF}}}$ and 
$\mathbf{W}_{\mathrm{BB}}[t]=[\mathbf{w}_{\mathrm{BB},1}[t],\ldots,\mathbf{w}_{\mathrm{BB},K}[t]]
\in\mathbb{C}^{M_{\mathrm{RF}}\times K}$ denote the analog and digital beamforming matrices, respectively. Then, the transmit signal is given by

\begin{equation}
\mathbf{X}[t]
=
\mathbf{W}_{\mathrm{RF}}[t]\mathbf{W}_{\mathrm{BB}}[t]\mathbf{x}[t].
\end{equation}

The transmit covariance matrix is expressed as
\begin{equation}
\mathbf{R}_x[t]
=
\mathbf{W}_{\mathrm{RF}}[t]\mathbf{W}_{\mathrm{BB}}[t]
\mathbf{W}_{\mathrm{BB}}^{\mathrm{H}}[t]\mathbf{W}_{\mathrm{RF}}^{\mathrm{H}}[t],
\end{equation}
and the transmit power constraint is
\begin{equation}
\operatorname{tr}(\mathbf{R}_x[t])\leq P_{\max}.
\end{equation}

% \begin{figure}[!t]
% 	\centering
% 	\includegraphics[
% 		width=0.72\linewidth,
% 		height=0.26\textheight,
% 		keepaspectratio
% 	]{system model.pdf}
% 	\caption{Integrated sensing and communication system model.}
% 	\label{fig-system-model}
% \end{figure}

\subsection{Communication Model}

Let $\mathbf{h}_k[t]\in\mathbb{C}^{M_t\times 1}$ denote the channel vector from the active AP to the $k$-th downlink vehicular user. Considering both multi-user downlink interference and cross-link interference from uplink users, the received SINR of the $k$-th vehicular user is given by
\begin{equation}
\mathrm{SINR}_k[t]
=
\frac{
\left|
\mathbf{h}_k^{\mathrm{H}}[t]\mathbf{W}_{\mathrm{RF}}[t]\mathbf{w}_{\mathrm{BB},k}[t]
\right|^2
}{
\sum\limits_{i\neq k}
\left|
\mathbf{h}_k^{\mathrm{H}}[t]\mathbf{W}_{\mathrm{RF}}[t]\mathbf{w}_{\mathrm{BB},i}[t]
\right|^2
+
p_u[t]\left|H_{k,u}[t]\right|^2
+
\sigma_k^2
},
\end{equation}
where $p_u[t]$ and $H_{k,u}[t]$ denote the transmit power and cross-link channel of the $u$-th uplink user, respectively, and $\sigma_k^2$ is the noise power.

With normalized bandwidth, the downlink spectral efficiency is
\begin{equation}
\mathrm{SE}^{\mathrm{DL}}_k[t]
=
\log_2\left(1+\mathrm{SINR}_k[t]\right).
\end{equation}

To characterize the reliability of short-packet vehicular communications, 
the finite blocklength normal approximation is adopted \cite{5452208}. 
Given the target downlink rate $R_{\mathrm{th}}^{\mathrm{DL}}$ and codeword length $L$, 
the outage probability of the $k$-th vehicular user is approximated as
\begin{equation}
P_{k}^{\mathrm{out}}[t]
=
Q\left(
\frac{
\mathrm{SE}^{\mathrm{DL}}_k[t]-R_{\mathrm{th}}^{\mathrm{DL}}
}{
\sqrt{V_k[t]/L}
}
\right),
\end{equation}
where $Q(\cdot)$ is the Gaussian $Q$-function, and $V_k[t]$ is the channel dispersion term, given by
\begin{equation}
V_k[t]
=
\left(
1-
\left(1+\mathrm{SINR}_k[t]\right)^{-2}
\right)
\left(\log_2 e\right)^2.
\end{equation}

The average system outage probability is defined as
\begin{equation}
\bar{P}_{\mathrm{sys}}^{\mathrm{out}}[t]
=
\frac{1}{K}
\sum_{k=1}^{K}
P_{k}^{\mathrm{out}}[t].
\end{equation}

\subsection{Sensing Model}

We consider a far-field point target, where the ISAC signal transmitted by the active AP is reflected by the target and received by multiple passive APs. For the $z$-th passive AP, the received echo depends on the transmit steering vector $\mathbf{b}_{t}(\theta_t)$, the receive steering vector $\mathbf{b}_{r,z}(\theta_{r,z})$, and the bistatic reflection coefficient $\beta_z$, where $\theta_t$ and $\theta_{r,z}$ denote the angle of departure and angle of arrival, respectively.

Based on the point-target angle estimation model, the Fisher information associated with $\theta_{r,z}$ is proportional to the target-direction transmit beam gain. For compact notation, it is written as
\begin{equation}
J_{z}[t]
=
\kappa_z
\mathbf{b}_{t}^{\mathrm{H}}(\theta_t)
\mathbf{R}_{x}[t]
\mathbf{b}_{t}(\theta_t),
\end{equation}
where $\kappa_z$ collects the effects of the reflection coefficient, receiver noise, sensing snapshots, and receive steering vector derivative.

Let $\mathbf{p}_e=[x_e,y_e]^{\mathrm{T}}$ and $\mathbf{p}_z=[x_z,y_z]^{\mathrm{T}}$ denote the target position and the $z$-th passive AP position, respectively. Denote the Jacobian vector between the target position and the arrival angle by $\mathbf{g}_z=\partial \theta_{r,z}/\partial \mathbf{p}_e$. Since the observations from different passive APs are independent, the Fisher information matrix for target localization is
\begin{equation}
\mathbf{F}_{p_e}[t]
=
\sum_{z\in\mathcal{Z}_{p}}
J_z[t]\mathbf{g}_z\mathbf{g}_z^{\mathrm{T}}.
\end{equation}
Accordingly, the position CRB is given by
\begin{equation}
\mathrm{CRB}(\mathbf{p}_e;t)
=
\operatorname{tr}
\left(
\mathbf{F}_{p_e}^{-1}[t]
\right).
\end{equation}

Therefore, the point-target sensing performance is determined by both the target-direction transmit beam gain and the spatial diversity of passive APs. By optimizing the transmit beamforming, the point-target CRB can be reduced to improve multistatic sensing accuracy.

\subsection{Problem Formulation}
We aim to optimize the hybrid beamforming matrices at the active AP to minimize the point-target CRB while guaranteeing vehicular communication reliability. The reliability-constrained multistatic ISAC beamforming problem is formulated as
\begin{equation}
\begin{aligned}
\mathcal{P}:
\quad
\min_{\mathbf{W}_{\mathrm{RF}}[t],\mathbf{W}_{\mathrm{BB}}[t]}
\quad
&
\mathrm{CRB}(\mathbf{p}_e;t)
\\
\mathrm{s.t.}
\quad
&
\bar{P}_{\mathrm{sys}}^{\mathrm{out}}[t]\leq P_{\mathrm{th}},
\\
&
\operatorname{tr}\left(\mathbf{R}_x[t]\right)\leq P_{\max},
\\
&
\left|[\mathbf{W}_{\mathrm{RF}}[t]]_{i,j}\right|=1,
\quad (i,j)\in\mathcal{W}_{\mathrm{RF}},
\end{aligned}
\end{equation}
where $P_{\mathrm{th}}$ denotes the maximum allowable average outage probability. The constraints respectively ensure communication reliability, limit the transmit power, and impose the constant-modulus requirement of the analog beamformer.

Problem $\mathcal{P}$ is non-convex due to the coupling among outage probability, transmit covariance, and hybrid beamforming variables. Therefore, a reliability-aware learning-based beamforming method is developed in the next section.

\section{Algorithm Design}
The formulated problem involves coupled continuous beamforming variables
and communication reliability constraints, which makes direct optimization
challenging. To enable efficient online beamforming under time-varying
channels, we recast it as a constrained Markov decision process and develop
a PI-LPPO algorithm.

\subsection{ The proposed algorithm}

In the formulated constrained Markov decision process, the system state contains the available communication
channel and target sensing information, while the action corresponds to the
beamforming variables generated by the policy network. The sensing objective
is represented by the reward function, whereas the communication outage
probability is incorporated into the cost function\cite{9539870}.

For a stochastic policy $\pi_{\theta}$, let
$J_r\left(\pi_{\theta}\right)$ and
$J_c\left(\pi_{\theta}\right)$ denote the expected cumulative sensing reward
and communication cost, respectively. The constrained policy optimization
problem is formulated as
\begin{equation}
\begin{aligned}
\max_{\theta}\quad
& J_r\left(\pi_{\theta}\right) \\
\mathrm{s.t.}\quad
& J_c\left(\pi_{\theta}\right)
\leq C_{\max},
\end{aligned}
\label{eq:cmdp_objective}
\end{equation}
where $C_{\max}$ denotes the prescribed communication cost threshold.

By introducing a nonnegative Lagrange multiplier $\lambda$, the constrained
problem is transformed into the following Lagrangian objective:
\begin{equation}
\mathcal{L}\left(\pi_{\theta},\lambda\right)
=
J_r\left(\pi_{\theta}\right)
-
\lambda
\left[
J_c\left(\pi_{\theta}\right)
-
C_{\max}
\right],
\label{eq:lagrangian_objective}
\end{equation}
where $\lambda \geq 0$ penalizes communication constraint violations.

Conventional Lagrangian PPO (LPPO) updates the multiplier according to the
observed constraint violation, which may cause delayed response or
oscillations near the constraint boundary. To improve constraint
adaptation, we introduce a PI-inspired multiplier $\lambda_m^{\mathrm{PI}}$ at the $m$-th policy update, which accounts for both the current constraint deviation and accumulated historical deviations.

Two critic networks are used to separately estimate the sensing reward and communication cost value functions. The corresponding advantages are combined as
\begin{equation}
\widehat{A}_t
=
\widehat{A}_t^{\,r}
-
\lambda_m^{\mathrm{PI}}
\widehat{A}_t^{\,c},
\label{eq:combined_advantage}
\end{equation}
where $\widehat{A}_t^{\,r}$ and $\widehat{A}_t^{\,c}$ denote the normalized
reward and cost advantages, respectively.

The probability ratio between the current and old policies is defined as

\begin{equation}
\rho_t\left(\theta\right)
=
\frac{
\pi_{\theta}\left(a_t\mid s_t\right)
}{
\pi_{\theta_{\mathrm{old}}}\left(a_t\mid s_t\right)
}.
\label{eq:probability_ratio}
\end{equation}

Based on the combined advantage, the policy network is updated using the clipped PPO surrogate objective:
\begin{equation}
\begin{aligned}
\mathcal{L}_{\mathrm{clip}}\left(\theta\right)
=
\mathbb{E}_t
\Bigg[
\min\Bigg(
& \rho_t\left(\theta\right)\widehat{A}_t, \\
& \operatorname{clip}\left(
\rho_t\left(\theta\right),
1-\epsilon_{\mathrm{clip}},
1+\epsilon_{\mathrm{clip}}
\right)
\widehat{A}_t
\Bigg)
\Bigg].
\end{aligned}
\label{eq:ppo_objective}
\end{equation}

The clipping operation limits overly large policy updates, while the
PI-adjusted multiplier dynamically balances sensing performance and
communication reliability.

\subsection{Algorithmic Procedure}

During training, the agent observes the current communication and sensing
state and generates a continuous beamforming action through the policy
network. The action is then projected to satisfy the constant-modulus
hardware constraint and normalized to meet the transmit power constraint.
The feasible beamforming solution is executed in the environment, which
returns the sensing reward, communication cost, and next state.

After collecting a batch of trajectories, the reward and cost critic
networks are updated to estimate the corresponding advantages. 
The effective Lagrange multiplier is adjusted according to the observed constraint
violation, and the policy network is updated using the clipped PPO objective.
This process is repeated until convergence.

During online execution, the active AP obtains a feasible beamforming
solution through a single forward pass of the trained policy network.
Thus, the nonconvex problem does not need to be solved repeatedly, which
reduces online computational complexity.

\section{ Simulation Analysis}

Simulation experiments are conducted to evaluate the convergence and reliability-constraint satisfaction of the proposed PI-LPPO algorithm. The system average outage probability is adopted as the communication cost, and the cost threshold is set to $C_{\max}=0.1$. For comparison, LPPO, PPO, SAC, multi-head double deep Q-network, and random policy are implemented under the same system model and simulation parameters.
\begin{figure}[!t]
	\centering
	\captionsetup[subfloat]{font=scriptsize}
	\subfloat[Reward]{
		\includegraphics[width=0.47\linewidth]{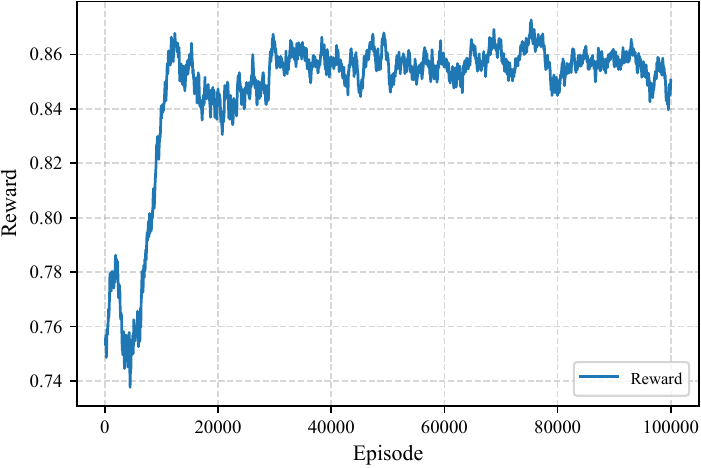}
		\label{fig-reward}
	}
	\hfill
	\subfloat[Cost]{
		\includegraphics[width=0.47\linewidth]{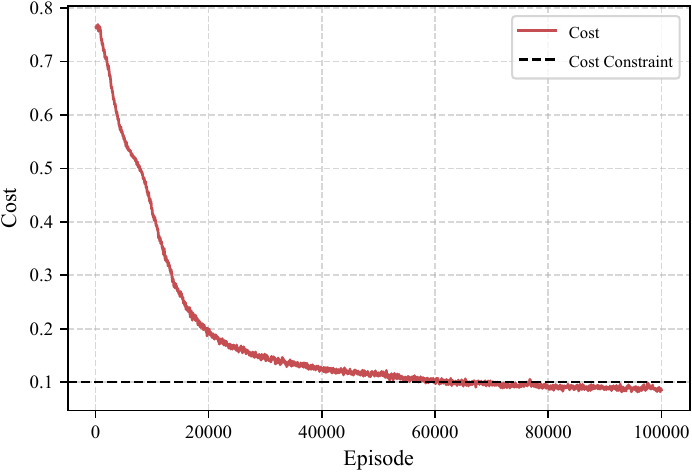}
		\label{fig-cost}
	}
	\caption{Convergence behavior of the proposed algorithm.}
	\label{fig-convergence}
\end{figure}

Fig.~1 (a) shows the convergence behavior of the sensing reward of the proposed PI-LPPO algorithm. The reward fluctuates in the early stage due to policy exploration, then increases rapidly and gradually stabilizes at around 0.85-0.86 after $2\times10^{4}$ episodes. This indicates that the agent learns a stable beamforming policy with favorable sensing performance.

Fig.~1 (b) shows the convergence behavior of the communication cost. The cost decreases from about 0.75 at the beginning of training and falls below the threshold of 0.1 after about $6\times10^{4}$ episodes, finally stabilizing at around 0.08--0.09. Together with Fig.~2, this result shows that PI-LPPO can reduce the outage cost while maintaining stable sensing performance.

\begin{figure}
	\centering
	\includegraphics[width = 0.78\linewidth]{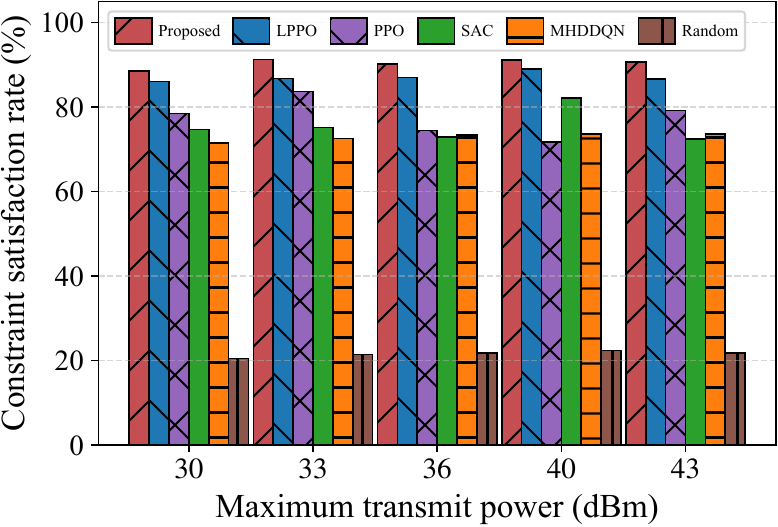}\\
	\caption{Communication constraint satisfaction rate vs. transmit power.}
	\label{fig-successrate}
\end{figure}

\begin{figure}
	\centering
	\includegraphics[width = 0.80\linewidth]{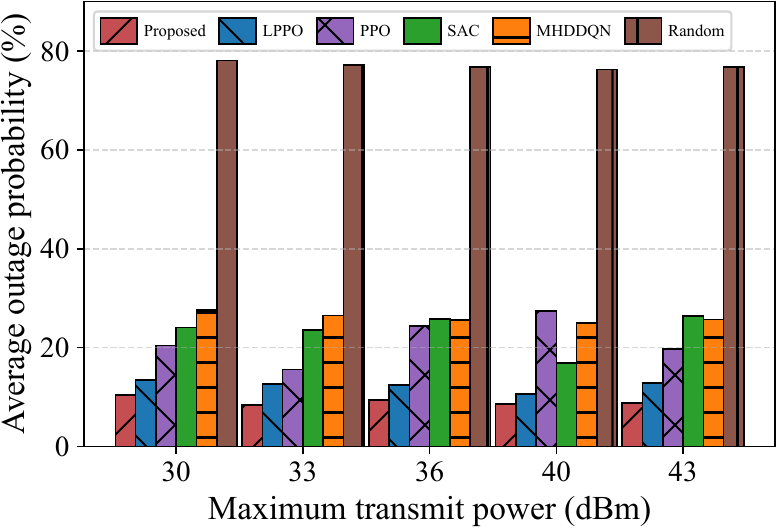}\\
	\caption{Average communication outage probability vs. transmit power.}
	\label{fig-outage-probability}
\end{figure}

Fig.~2 compares the communication reliability constraint satisfaction rates under different maximum transmit powers. PI-LPPO achieves the highest and most stable satisfaction rate among all schemes. This is because the proportional component enables fast responses to current constraint violations, while the integral component helps suppress long-term accumulated violations.

Fig.~3 compares the average outage probabilities of different algorithms. PI-LPPO keeps the outage probability at about 8\%--10\% and satisfies the prescribed reliability threshold, while other learning-based baselines show higher outage probabilities or occasional constraint violations. These results verify the effectiveness of the proposed PI-Lagrangian mechanism in improving reliability constraint control.

\section{ Conclusion}

This letter studied reliability-constrained hybrid beamforming for a transceiver-separated multistatic ISAC system in vehicular networks. A target-position CRB minimization problem was formulated under communication reliability, transmit power, and analog constant-modulus constraints. To solve this non-convex problem online, it was reformulated as a CMDP, and a PI-LPPO-based beamforming method was developed. 
Simulation results showed that the proposed algorithm achieves stable convergence, keeps the average outage probability below the reliability threshold, and outperforms baseline methods in outage probability and reliability-constraint satisfaction.

\bibliographystyle{IEEEtran}
\bibliography{ref}
\vfill
\end{document}